\documentclass[journal, 10pt]{IEEEtran}
\usepackage[tbtags]{amsmath}
\usepackage{graphicx}
\usepackage{amssymb}
\usepackage{bm}
\usepackage{multirow}
\usepackage{adjustbox}
\usepackage{tabularx}
\usepackage{tikz}
\usepackage{caption}
\usepackage{subcaption}
\usepackage{verbatim}
\usepackage{float}
\usepackage{hyperref}
\usepackage{xcolor}
\usepackage{colortbl}
\usepackage{soul}

\begin{document}

\title{On Cross-Corpus Generalization of Deep Learning Based Speech Enhancement}

\author{Ashutosh~Pandey, ~\IEEEmembership{Student~Member,~IEEE}~and~DeLiang~Wang,~\IEEEmembership{Fellow,~IEEE}

\thanks{This research was supported in part by two NIDCD grants (R01DC012048 and R01DC015521) and the Ohio Supercomputer
Center.}
\thanks{A. Pandey is with the Department of Computer Science and Engineering, The Ohio State University, Columbus, OH 43210 USA (e-mail:
pandey.99@osu.edu).}
\thanks{D. L. Wang is with the Department of Computer Science and Engineering
and the Center for Cognitive and Brain Sciences, The Ohio State University,
Columbus, OH 43210 USA (e-mail: dwang@cse.ohio-state.edu)}\vspace{-0.5em}}

\maketitle

\begin{abstract}
In recent years, supervised approaches using deep neural networks (DNNs) have become the mainstream for speech enhancement. It has been established that DNNs generalize well to untrained noises and speakers if trained using a large number of noises and speakers. However, we find that DNNs fail to generalize to new speech corpora in low signal-to-noise ratio (SNR) conditions. In this work, we establish that the lack of generalization is mainly due to the channel mismatch, i.e. different recording conditions between the trained and untrained corpus. Additionally, we observe that traditional channel normalization techniques are not effective in improving cross-corpus generalization. Further, we evaluate publicly available datasets that are promising for generalization. We find one particular corpus to be significantly better than others. Finally, we find that using a smaller frame shift in short-time processing of speech can significantly improve cross-corpus generalization. The proposed techniques to address cross-corpus generalization include channel normalization, better training corpus, and smaller frame shift in short-time Fourier transform (STFT). These techniques together improve the objective intelligibility and quality scores on untrained corpora significantly. 
\end{abstract}


\IEEEpeerreviewmaketitle

\section{Introduction}
Speech signal in a real-world environment is degraded by background noise. A degraded speech signal can severely degrade the performance of speech-based applications such as automatic speech recognition (ASR), speaker identification, and hearing aids. Speech enhancement is concerned with improving the intelligibility and quality of a speech signal degraded by additive noise, and commonly used as preprocessors in speech-based applications to improve their performance in noisy environments.

In real-world environments, speech signals are varied or distorted \cite{benzeghiba2007automatic}. Sources of variations include background noise, room reverberation, speaker, language, accent, and communication channel. Ideally a speech enhancement algorithm should work well in different acoustic conditions. However, developing a general algorithm that works in all conditions remains a technical challenge. 

Traditional approaches to speech enhancement include spectral subtraction \cite{boll1979suppression}, Wiener filtering \cite{scalart1996speech}, statistical model-based methods \cite{loizou2013speech}, and nonnegative matrix factorization \cite{mohammadiha2013supervised}. These approaches work well for stationary noises but have difficulty in handling nonstationary noises or a large number of speakers. In recent years, deep learning-based approaches have become the mainstream for speech enhancement (see \cite{wang2017supervised} for an overview). Among the most popular deep learning approaches are fully-connected networks \cite{wang2013towards, xu2015regression}, recurrent neural networks (RNNs) \cite{weninger2015speech, chen2017long} and convolutional neural networks (CNNs) \cite{fu2016snr, tan2018gated1, pandey2019new}.

In \cite{chen2016large}, Chen \textit{et al.} demonstrated that fully connected feedforward networks trained for a single speaker, using a large number of noises, can generalize to untrained noises. However, such a network has difficulty generalizing to both of untrained speakers and noises, when trained using a large number of noises and speakers \cite{chen2017long}. In \cite{chen2017long}, a RNN with long short-term memory (LSTM) is employed to develop a speaker- and noise-independent model for speech enhancement. This was achieved by training a four-layered RNN model using utterances from 77 speakers mixed with 10000 different noises.

In the last few years, speech enhancement research has aimed to improve the performance of speaker-and noise-independent models. In \cite{tan2018gated1}, the authors propose a CNN with gated and dilated convolutions for magnitude-spectrum enhancement. A recent trend is the enhancement of phase, obtaining better speech enhancement than the magnitude-only enhancement approaches. The two popular approaches are complex-spectrogram enhancement \cite{williamson2016complex, tan2019learning, fu2017complex, pandey2019exploring, choi2019phase} and time-domain enhancement \cite{fu2017raw, pandey2019new, pascual2017segan, qian2017speech, rethage2017wavenet, pandey2019tcnn}.

The common practice in all the DNN based approaches is that a DNN is trained using utterances of different speakers from a single corpus and evaluated on untrained speakers from the same corpus. However, we find that when evaluated on utterances from untrained corpora, DNN performance may degrade significantly. This behavior has not been revealed and analyzed before. To be suitable for real-world applications, speech enhancement has to work on noisy utterances recorded in an unknown fashion, i.e. on any untrained corpus.

In this study, we perform an experimental study to understand cross-corpus generalization of DNNs. Our key observation is that the generalization gap is severe at low SNR conditions and is mainly due to the channel mismatch between different speech corpora. We examine the effectiveness of traditional channel normalization techniques for speech enhancement in low SNR conditions. 

The general behavior of traditional channel normalization methods used in ASR or speaker identification systems, such as cepstral mean subtraction (CMS) \cite{atal1976automatic, furui1981cepstral} or RASTA filtering \cite{hermansky1991compensation, hermansky1994rasta}, is unknown for supervised speech enhancement. In supervised approaches to speech enhancement, a noisy utterance is generated by adding a noise segment to a clean speech utterance. It is highly unlikely that the channels of clean speech and noise will be similar. This creates a channel situation that is different from those in ASR and speaker recognition where the noise channel is not a main concern.  In other words, a noisy utterance captures two kinds of channel effects, one for speech and the other for noise. This implies that the predicted channel from the noisy utterance may be inaccurate in noise dominant segments. To verify this analysis, we have evaluated two different channel normalization methods, mean subtraction and RASTA filtering in the log-spectrum domain. We choose the log-spectrum domain because most of the DNN based speech enhancement systems use either spectrum or log-spectrum as the input features. We observe improved enhancement using channel normalization, however, the improvements are indeed limited in low SNR conditions. 

Further, we evaluate different corpora that are promising for cross-corpus generalization. A corpus that is recorded using many microphones or recorded in different acoustic conditions would be promising as it will expose the underlying DNN model to different channels. LibriSpeech \cite{panayotov2015librispeech} and VoxCeleb2 \cite{Chung18b} are two such corpora. The utterances in LibriSpeech are extracted from audiobooks that are read by different volunteers across the globe. This implies that the utterances recorded by different volunteers have different channel characteristics. VoxCeleb2 utterances are extracted from the audios in YouTube videos and hence are recorded in different conditions and using different devices. We find LibriSpeech to be significantly better than VoxCeleb2 and WSJ \cite{paul1992design}, the latter commonly used in speaker-independent enhancement models. 

Additionally, we investigate the use of smaller frame 	shifts in STFT, as smaller shifts may lead to better cross-corpus generalization because of the averaging effect in the overlap-and-add stage of inverse STFT. This turns out to be a very simple and effective technique for improving cross-corpus generalization.

Finally, we combine all the proposed techniques; channel normalization, better training corpus, and smaller frame shift. This combination substantially improves objective intelligibility and quality scores. The short-time objective intelligibility (STOI) \cite{taal2011algorithm} and the perceptual evaluation of speech quality (PESQ) \cite{rix2001perceptual} scores at $-5$ dB SNR for babble noise are improved by $13.9$ percentage points and $0.59$ respectively for the utterances of a male speaker in the challenging IEEE corpus \cite{rothauser1969ieee}.

To our knowledge, this is the first systematic study on cross-corpus generalization in DNN based speech enhancement. The results of this study, we believe, represent a major step towards robust speech enhancement in real-world conditions. The rest of the paper is organized as follows. In Section II, we describe the speech enhancement framework used in this study. Section III explains corpus channel. Section IV illustrates the corpus fitting problem in speech enhancement. In Section V, we describe the techniques explored in this study to improve cross-corpus generalization. Experimental settings are given in Section VI and Section VII presents the results. Concluding remarks are given in Section VIII. 
\section{Deep Learning based speech enhancement}
\subsection{Problem Definition}
Given a clean speech signal $\bm{x}$ and a noise signal $\bm{n}$, the noisy speech signal is formed by the additive mixing as the following
\begin{equation}
\bm{y} = \bm{x} + \bm{n}
\end{equation}
where \{$\bm{y}$, $\bm{x}$, $\bm{n}$\} $ \in \mathbb{R}^{M \times 1}$. $M$ represents the number of samples in the signal. The goal of a speech enhancement algorithm is to get a close estimate, $\widehat{\bm{x}}$,  of $\bm{x}$ given $\bm{y}$.
\subsection{Data Generation}

Given a speech corpus $\bm{C}$ containing $N_{tr}$ training utterances \{$\bm{x}_{tr}^{1}, \bm{x}_{tr}^{2}, ..., \bm{x}_{tr}^{N_{tr}}$\} and $N_{te}$ test utterances \{$\bm{x}_{te}^{1}, \bm{x}_{te}^{2}, ..., \bm{x}_{te}^{N_{te}}$\}, we denote $\bm{C}_{tr}$ as the set of training utterances and $\bm{C}_{te}$ as the set of test utterances in corpus $\bm{C}$.

The noisy utterances are generated by artificially adding noises to the utterances in $\bm{C}_{tr}$ and $\bm{C}_{te}$.
\begin{align}
\bm{y}_{tr}^{i} &= \bm{x}_{tr}^{i} + \bm{n}_{tr}^{i}, \ \ \ \ i = 1, 2, ... N^{tr}\\
\bm{y}_{te}^{j} &=  \bm{x}_{te}^{j} + \bm{n}_{te}^{j}, \ \ \ \ j = 1, 2, ... N^{te}
\end{align}
In general, to assess noise generalization, $\bm{n}_{tr}^{i}$ and $\bm{n}_{te}^{j}$ are set to be either different noises or different segments of nonstationary noises. Similarly, to assess speaker generalization, speakers in $\bm{C}_{tr}$ and $\bm{C}_{te}$ are set to be different.

In this work, we evaluate DNN based speech enhancement models for cross-corpus generalization. We train different models on corpora \{$\bm{C}^{1}_{tr}$, $\bm{C}^{2}_{tr}$, ..., $\bm{C}^{P_{tr}}_{tr}$\} but evaluate them on utterances from untrained corpora  \{$\hat{\bm{C}}^{1}_{te}$, $\hat{\bm{C}}^{2}_{te}$, ..., $\hat{\bm{C}}^{P_{te}}_{te}$\}. $P_{tr}$ and $P_{te}$ denote the numbers of training and test corpora respectively.

\subsection{Feature Extraction and Training Targets}
The pairs \{$\bm{x}$,  $\bm{y}$, $\bm{n}$\} are transformed to the time-frequency (T-F) representation using STFT. 
\begin{align}
\bm{X} &= \text{STFT}(\bm{x})\\
\bm{Y} &= \text{STFT}(\bm{y})\\
\bm{N} &= \text{STFT}(\bm{n})
\end{align} 
where \{$\bm{X}$, $\bm{Y}$, $\bm{N}$\} $\in$ $\mathbb{C}^{T \times F}$, and $T$ and $F$ represent the number of frames and number of frequency bins. In this study, we use either STFT magnitude $|\bm{Y}|$ or logarithm of STFT magnitude, log$|\bm{Y}|$, as the input feature.

There are many training targets studied in the literature such as the ideal ratio mask (IRM) \cite{wang2014training}, STFT magnitude \cite{xu2015regression}, and spectral magnitude mask (SMM) \cite{wang2014training}. We use the IRM in this study,  defined as:
\begin{equation}
IRM(t, f) = \sqrt{\frac{|X(t, f)|^{2}}{|X(t, f)|^{2} + |N(t, f)|^{2}}}
\end{equation}
where $X(t, f)$,  $N(t, f)$ and $IRM(t, f)$, respectively, denote the values of $\bm{X}$, $\bm{N}$ and $\bm{IRM}$ at the corresponding T-F unit.

\subsection{Model Architecture}
We use a 4-layer bidirectional LSTM (BLSTM) network with 512 hidden units in each direction. One fully-connected layer with 512 units is used before the BLSTM, which is followed by a fully-connected layer at the output with sigmoidal nonlinearity.

\subsection{Loss Function}
The BLSTM network takes as input the feature, $|\bm{Y}|$ or  log$|\bm{Y}|$, and outputs the estimated IRM, $\bm{RM}$. A mean squared error (MSE) loss is used between $\bm{IRM}$ and $\bm{RM}$. The utterance level MSE loss is given below.
\begin{equation}
L = \frac{1}{TF}\sum_{t=0}^{T} \sum_{f=0}^{F}[IRM(t, f) - RM(t, f)]^2
\end{equation}

\subsection{Time Domain Reconstruction}
The trained model is used for predicting the IRM of noisy utterances in the test set. $\bm{RM}$ is multiplied to the noisy STFT magnitude, $|\bm{Y}|$, to obtain the enhanced STFT magnitude, $|\widehat{\bm{X}}|$. 
\begin{equation}
|\widehat{\bm{X}}| = |\bm{Y}| \otimes \bm{RM}
\end{equation}
where $\otimes$ denotes element-wise multiplication.

The estimated STFT magnitude is combined with the noisy STFT phase to obtain the estimated STFT.
\begin{equation}
\widehat{\bm{X}} =|\widehat{\bm{X}}| \otimes e^{j\angle \bm{Y}}
\end{equation}
where $\angle \bm{Y}$ represents the noisy phase.
Finally, inverse STFT is used to obtain the enhanced waveform.
\begin{equation}
\widehat{\bm{x}} = \text{ISTFT}(\widehat{\bm{X}})
\end{equation}

\section{Corpus Channel}
A speech corpus generally contains different utterances spoken by many speakers. The utterances are recorded in a controlled environment so that the recording is clean and suitable to be used for speech-based applications. The different controlled environments used for different corpora may lead to different stationary components in the utterances. For example, if recording microphones are different, a sentence spoken by the same person can be very different in quality. We refer to the stationary component of a corpus as the corpus channel. 

An algorithm developed and shown to be effective for one corpus may not work when evaluated on a corpus recorded in a different condition. To illustrate this, Fig. 1 plots the log-spectrum of an utterance from the TIMIT corpus \cite{garofolo1993darpa} that is convolved with two different microphone impulse response (MIR) functions\footnote{The two MIRs are obtained from \url{https://www. audiothing.net/impulses/vintage-mics/}}. We can observe that the energy patterns in the two spectra are very different. The left spectrum has higher energy around $100^{th}$ frequency bin and lower energy around the $0^{th}$ bin compared to the right spectrum. This type of difference in distribution may cause an algorithm to degrade on untrained corpora.
\begin{figure}[!h]
\centering
 \includegraphics[width=0.98\columnwidth, keepaspectratio]{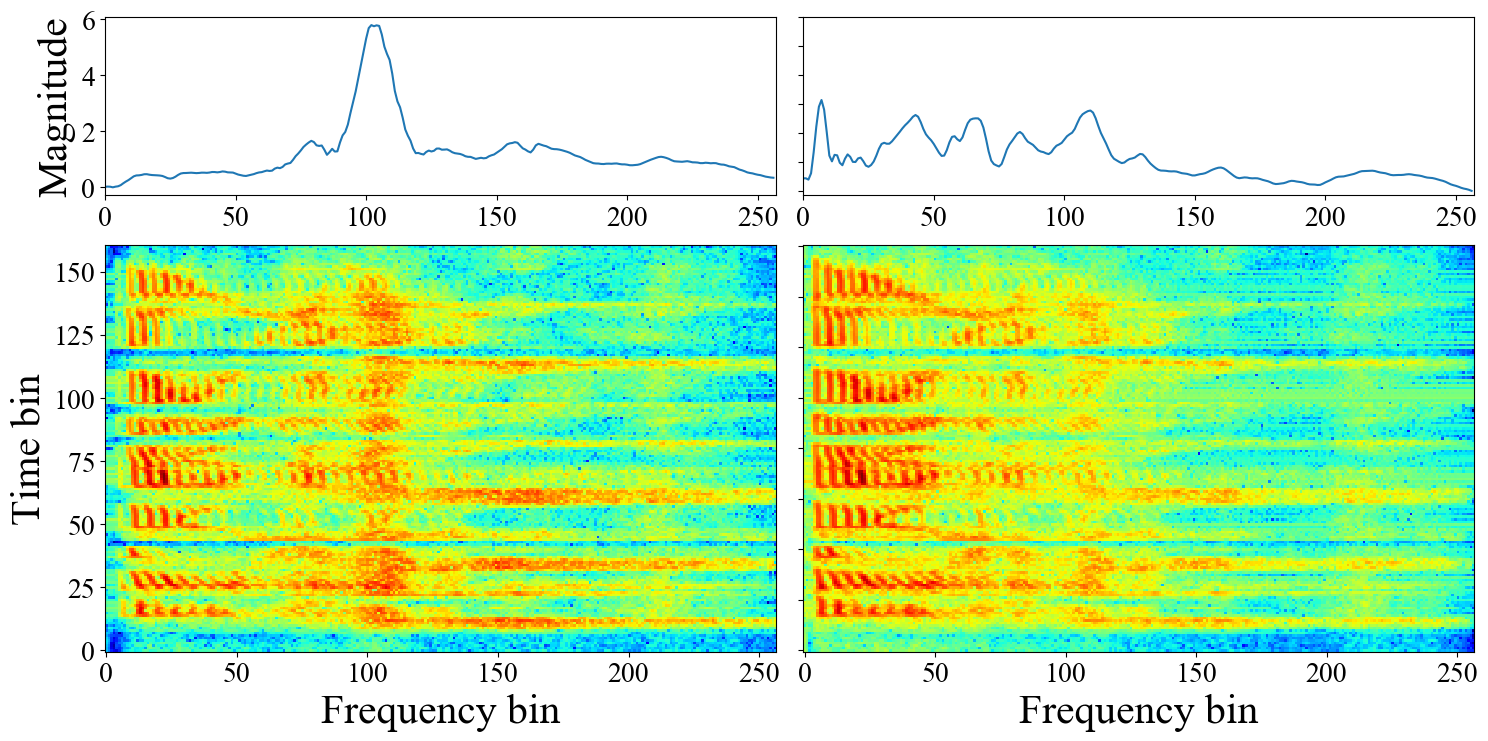}
\caption{Differences in the energy distribution of a spectrum convolved using different MIR functions. The frequency responses of MIRs are shown in the top row.}
\end{figure}
A stationary channel can be defined as a linear- and time-invariant filter given in the following equation,
\begin{equation}
\bm{x}= \bm{s} * \bm{h} = \sum_{k=0}^{K-1}s[n-k] \cdot h[k]
\end{equation}  
where $*$ denotes the convolution operator, $\bm{x}$ and $\bm{s}$ are discrete signals indexed by $n$, and $\bm{h}$ is a digital filter with $K$ taps. When the underlying signal, $\bm{s}$, is a time-varying speech signal, Equation 12 can be transformed into the following form using STFT. 
\begin{equation}
X(t, f) = S(t, f) \cdot H(f)
\end{equation} 
where $\bm{H}$ is the time-invariant but frequency-dependent gain introduced by the channel. Note that $H(f)$ does not contain any time index implying the stationarity of the channel. Taking the logarithm of complex magnitude in both sides of Equation 13, we get
\begin{equation}
\text{log}|X(t, f)| = \text{log}|S(t, f)| + \text{log}|H(f)|
\end{equation}
A straightforward method to remove stationary channel from a speech signal is log-spectral mean subtraction (LSMS). In this method, the long-term average of a log-spectrum is subtracted from the log-spectrum to obtain a channel removed log-spectrum.  Taking the average over time in Equation 14, we get
\begin{equation}
\frac{1}{T} \cdot \sum_{t}\text{log}|X(t, f)| = \frac{1}{T} \cdot \sum_{t}\text{log}|S(t, f)| + \text{log}|H(f)|
\end{equation}
Now, we define the channel of a corpus, $\bm{V}$, using the following equation.
\begin{equation}
\begin{split}
\text{log}|V(f)| &= \frac{\sum_{i=1}^{N_{tr}}\sum_{t=1}^{T}\text{log}|X_{tr}^{i}(t, f)|}{N_{tr} \cdot T} \\
&= \frac{\sum_{i=1}^{N_{tr}}\sum_{t=1}^{t=T}[\text{log}|S_{tr}^{i}(t, f)| + \text{log}|H(f)|]}{N_{tr} \cdot T}\\
&= \text{log}|\bar{S}(f)| + \text{log}|H(f)|
\end{split}
\end{equation}
Thus the defined corpus channel consists of two components, where $\bm{H}$ corresponds to the recording channel and  $\bm{\bar{S}}$ corresponds to the log-inverse of the average of log-spectrum over the corpus. It is important to note that channel differences between corpora are primarily caused by $\bm{H}$, as the long-term average speech spectrum is similar across different dialects of the same language and even different languages \cite{byrne1994international}.

Further subtracting Equation 16 from Equation 14, we get
\begin{equation}
\text{log}|X(t, f)| - \text{log}|V(f)| = \text{log}|S(t, f)| - \text{log}|\bar{S}(f)|
\end{equation}
The above equation says that removing the defined corpus channel from an utterance of a corpus gives a normalized utterance with both channel and speech mean effects removed.

We will use Equation 16 to estimate the spectral magnitudes of the corpus channel of three popular corpora utilized for speech enhancement; WSJ SI-84, TIMIT, and IEEE \cite{rothauser1969ieee}. A frame of 20 ms with a shift of 10 ms is used for STFT computation. The estimates for the channels are plotted in Fig. 2. We can observe that the channels are quite different from each other. Even though the peaks occur at nearby frequencies, the decay rates are much different. The decay rate is fastest for IEEE and slowest for TIMIT. TIMIT and WSJ exhibit 2 peaks whereas IEEE shows only one peak.
\begin{figure}[!t]
\centering
 \includegraphics[width=0.95\columnwidth, keepaspectratio]{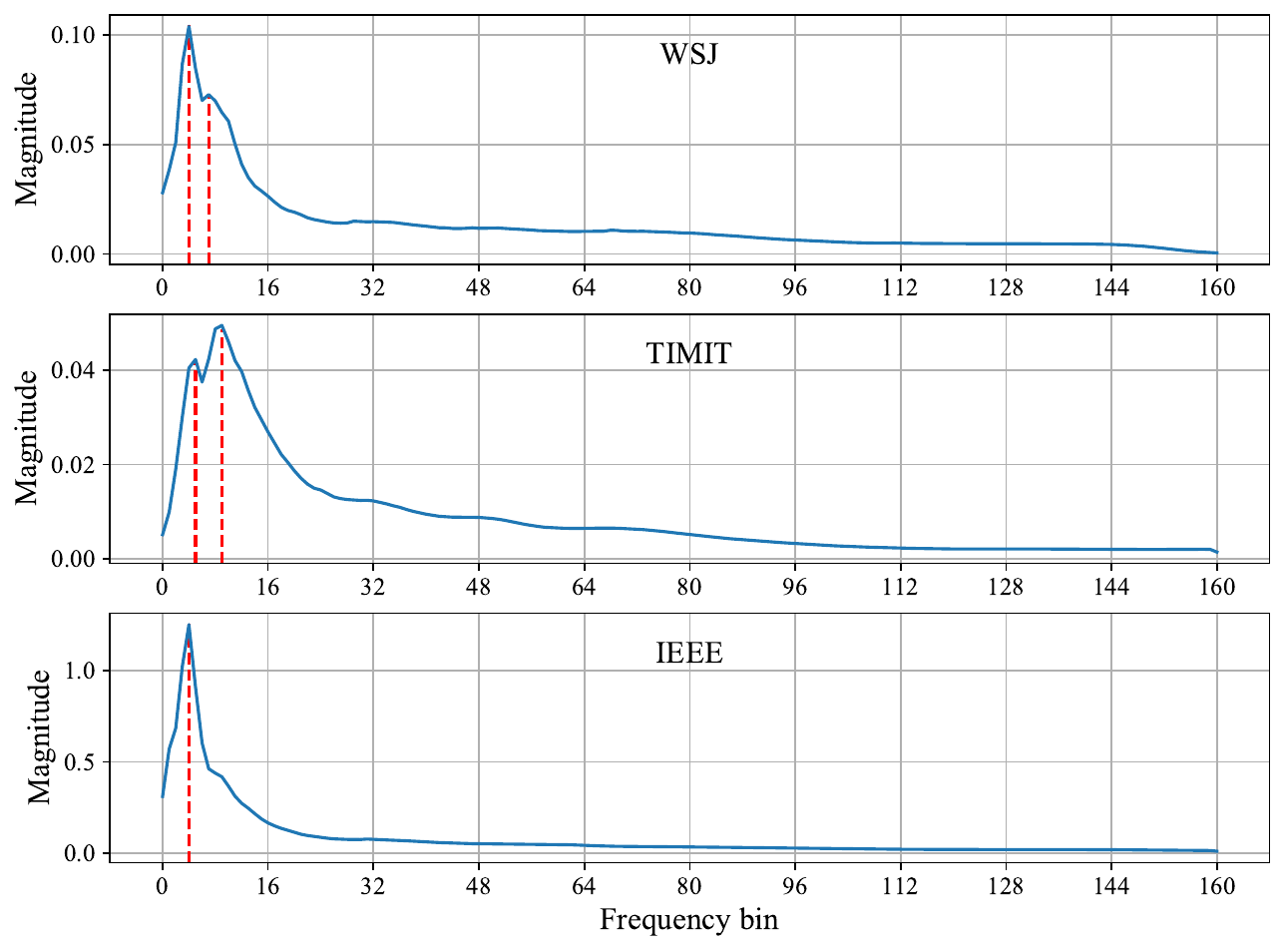}
		\caption{The estimated spectral magnitudes of the channels of three speech corpora.}
\end{figure}

\section{Corpus Fitting}
In this section, we demonstrate that models trained on one corpus fail to generalize to untrained corpora. Further, we show that the corpus channel is one of the factors that reduce the performance on untrained corpora. 

We evaluate three different types of models; an IRM based BLSTM model described in Section II, a complex-spectrum based model proposed in \cite{tan2019complex} and two time-domain models proposed in \cite{pandey2019new, pandey2019tcnn}. The models are trained on the WSJ corpus and are evaluated on 3 different corpora: WSJ, TIMIT, and IEEE. These corpora have been widely utilized in deep learning based speech enhancement studies. IEEE has a large number of utterances but few speakers, and is commonly used to train speaker-dependent models by using utterances of a single speaker \cite{chen2016large, williamson2016complex}; TIMIT has been used for small-scale training of noise-dependent and noise-independent models \cite{wang2014training, fu2016snr, fu2017raw, pandey2018adversarial, pandey2019exploring}, and WSJ has been used to train speaker- and noise-independent models \cite{chen2017long, tan2018gated1, pandey2019new, tan2019learning}. We select one male and one female speaker from  IEEE and treat them as two different corpora. They are denoted as IEEE Male and IEEE Female respectively. A detailed description of test data preparation is given in Section VI-A. The evaluation results in terms of STOI (\%) and PESQ, for babble noise at SNRs of $-5$ dB and $-2$ dB, are given in Table I. 

One can observe that the performance on the trained corpus, WSJ, is excellent. STOI is improved by more than 19.5\% for all the models. However, the improvements are much reduced on untrained corpora, TIMIT, IEEE Male and IEEE Female. For the IEEE Male speaker, AECNN-SM and CRN even degrade STOI compared to unprocessed mixtures. Similarly, PESQ is also degraded in many cases. The results suggest that the BLSTM model is better in terms of generalization, even though within-corpus enhancement results are not as good as the more recent models. Therefore we choose this model for comparisons in the rest of the paper.
\begin{table}[!b]
\centering
\caption{STOI and PESQ comparisons between different test corpora for four deep learning based speech enhancement methods.}
\centering
\begin{adjustbox}{width=0.95\columnwidth}
\begin{tabular}{|c|c|cc|cc|cc|cc|}
\hline
\multicolumn{2}{|c|}{ Test Corpus } & \multicolumn{2}{c|}{ WSJ } & \multicolumn{2}{c|}{ TIMIT } & \multicolumn{2}{c|}{ IEEE Male } & \multicolumn{2}{c|}{ IEEE Female } \\
\hline
\multicolumn{2}{|c|}{ Test SNR } & -5 dB & -2 dB & -5 dB & -2 dB & -5 dB & -2 dB & -5 dB & -2 dB \\
\hline
\multirow{5}{*}{ \rotatebox{90}{STOI (\%) }} & Mixture & 58.6 & 65.5 & 54.0 & 60.9 & 55.0 & 62.3 & 55.5 & 62.9 \\
\cline{2-10}
& BLSTM & 77.4 & 83.0 & 64.7 & 73.3 & 60.4 & 74.0 & 62.5 & 73.5 \\
& CRN \cite{tan2019complex}& 80.3 & 86.8 & 59.0 & 69.6 & 52.6 & 65.5 & 51.6 & 68.0 \\
& AECNN-SM \cite{pandey2019new}& 81.0 & 88.3 & 60.8 & 72.0 & 51.5 & 65.2 & 61.1 & 75.8 \\
& TCNN \cite{pandey2019tcnn}& 82.7 & 88.9 & 61.6 & 72.9 & 57.2 & 69.9 & 56.5 & 74.1 \\
\hline
\multirow{5}{*}{  \rotatebox{90}{PESQ} } & Mixture & 1.54 & 1.69 & 1.46 & 1.63 & 1.46 & 1.63 & 1.12 & 1.32 \\
\cline{2-10}
& BLSTM & 1.97 & 2.22 & 1.70 & 2.00 & 1.52 & 1.89 & 1.26 & 1.66 \\
& CRN \cite{tan2019complex}& 2.17 & 2.50 & 1.33 & 1.73 & 1.07 & 1.50 & 0.91 & 1.50 \\
& AECNN-SM \cite{pandey2019new}& 2.19 & 2.60 & 1.40 & 1.78 & 1.13 & 1.50 & 1.28 & 1.83 \\
& TCNN \cite{pandey2019tcnn}& 2.19 & 2.53 & 1.33 & 1.74 & 1.18 & 1.61 & 1.01 & 1.64 \\
\hline
\end{tabular}
\end{adjustbox}
\end{table}
\begin{figure*}[!t]
\centering
\includegraphics[width=0.75\textwidth]{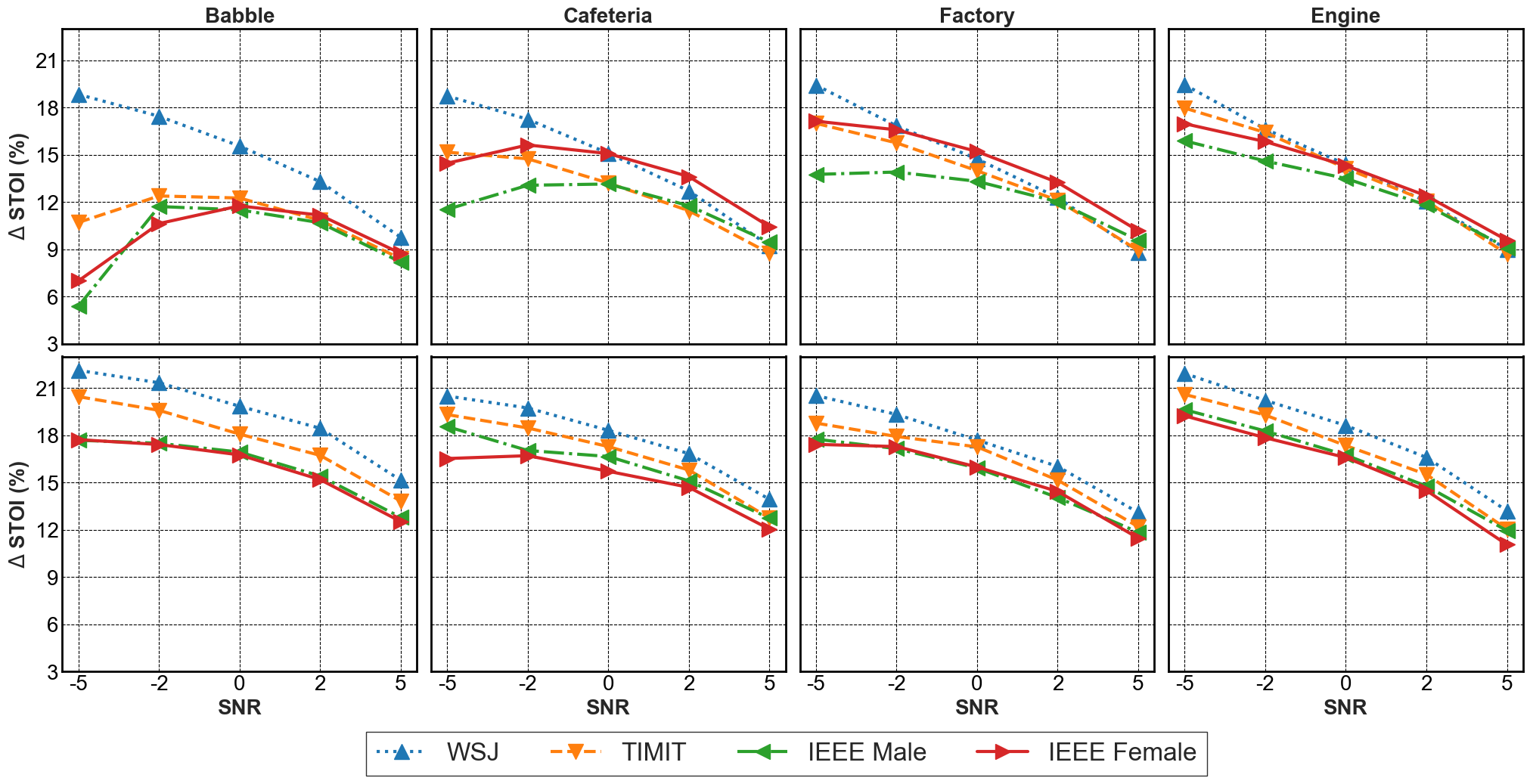}
\caption{Effects of corpus-channel on cross-corpus generalization. First row plots $\Delta$STOI (\%) obtained using original WSJ utterances. Second row plots $\Delta$STOI (\%) using channel-removed utterances.}
\end{figure*}

Next, we illustrate the behavior of the BLSTM model for different types of noises and at different SNR conditions. The plots of STOI improvement (\%) are shown in the first row of Fig. 3. We observe that for all the noises the gap between trained and untrained corpus is the largest at $-5$ dB and gradually narrows with increasing SNR. This illustrates that cross-corpus generalization is a severe issue in low SNR conditions. Similarly, the generalization gap at low SNRs for different noises is in order of babble, cafeteria, factory and engine.

Finally, we design an experiment to demonstrate that the corpus channel is a major culprit for the cross-corpus generalization issue. We use Equation 17 to get corpus channel removed spectrum of utterances in a corpus. The corpus channel removed spectrum is used for time-domain reconstruction using Eqs. 10 and 11. For a given corpus $\bm{C}$, we use $\bm{C}_{tr}$ for the corpus channel estimation, and use it to get corpus channel removed utterances in both $\bm{C}_{tr}$ and $\bm{C}_{te}$. We use a frame size of 2048 and frame shift of 32 in STFT. We find that this setting introduces negligible artifacts in the modified utterances.

We show the effect of corpus channel normalization on sample utterances from different corpora in Fig. 4. One can observe that the energy distribution in different frequency bins becomes more prominent, especially in the high-frequency range where the corpus channel has a large attenuation factor.

We use corpus channel normalized utterances to generate a new training corpus on WSJ and new test corpora on WSJ, TIMIT, IEEE Male and IEEE Female. The BLSTM model is trained on the new WSJ corpus and evaluated on all the test corpora for four different noises. The improvements in STOI (\%) are plotted in the second row of Fig. 3. These improvements are significantly higher than those in the first row. For example, $\Delta$STOI of the babble noise at $-5$ dB changes from $5\%$ to $18\%$ for IEEE Male, and $7\%$ to $18\%$ for IEEE Female. In addition, $\Delta$STOI improves for all the noises and in all SNR conditions. This demonstrates that the corpus channel is one of the main causes for the cross-corpus generalization issue, and channel differences need to be accounted for in order to improve cross-corpus generalization.

\section{Improving cross-corpus generalization}
In this section, we describe different techniques investigated in this study to improve cross-corpus generalization.

\subsection{Modified Loss Function} 
We find that using a loss over high energy T-F units is better for cross-corpus generalization. We use loss over T-F units within the 20 dB of the maximum amplitude T-F unit. A similar loss function has been utilized in speaker separation methods, such as deep clustering \cite{hershey2016deep}. The modified utterance level loss is given as 
\begin{equation}
L = \frac{\sum \limits_{t=0}^{T} \sum \limits_{f=0}^{F}[IRM(t, f) - RM(t, f)]^2 \cdot M(t, f)}{\sum \limits_{t=0}^{T} \sum \limits_{f=0}^{F}M(t, f)}
\end{equation}
where,
\begin{equation}
M(t, f) = \begin{cases}
1, & \ |Y(t, f)| \geq 0.01\cdot \text{Max($\bm{|Y|}$)} \\
0 & \ \text{Otherwise} 
\end{cases}
\end{equation}

\begin{figure}[!b]
\centering
\includegraphics[width=0.8\columnwidth]{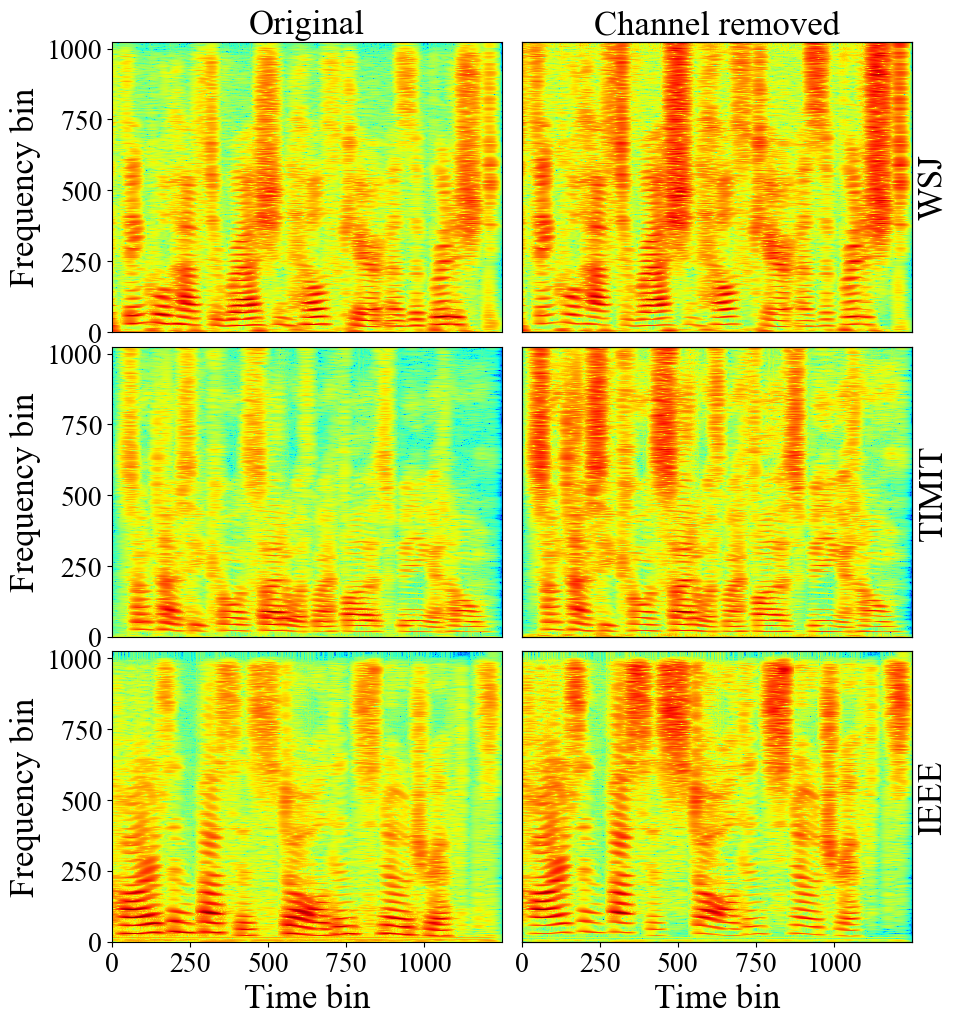}

\caption{Effects of channel normalization. The spectrogram of one utterance from each of the three corpora are plotted in the first column. The corresponding channel removed spectrograms are plotted in the second column. }
\end{figure}

 \begin{figure*}[!t]
\centering
\includegraphics[width=0.75\textwidth]{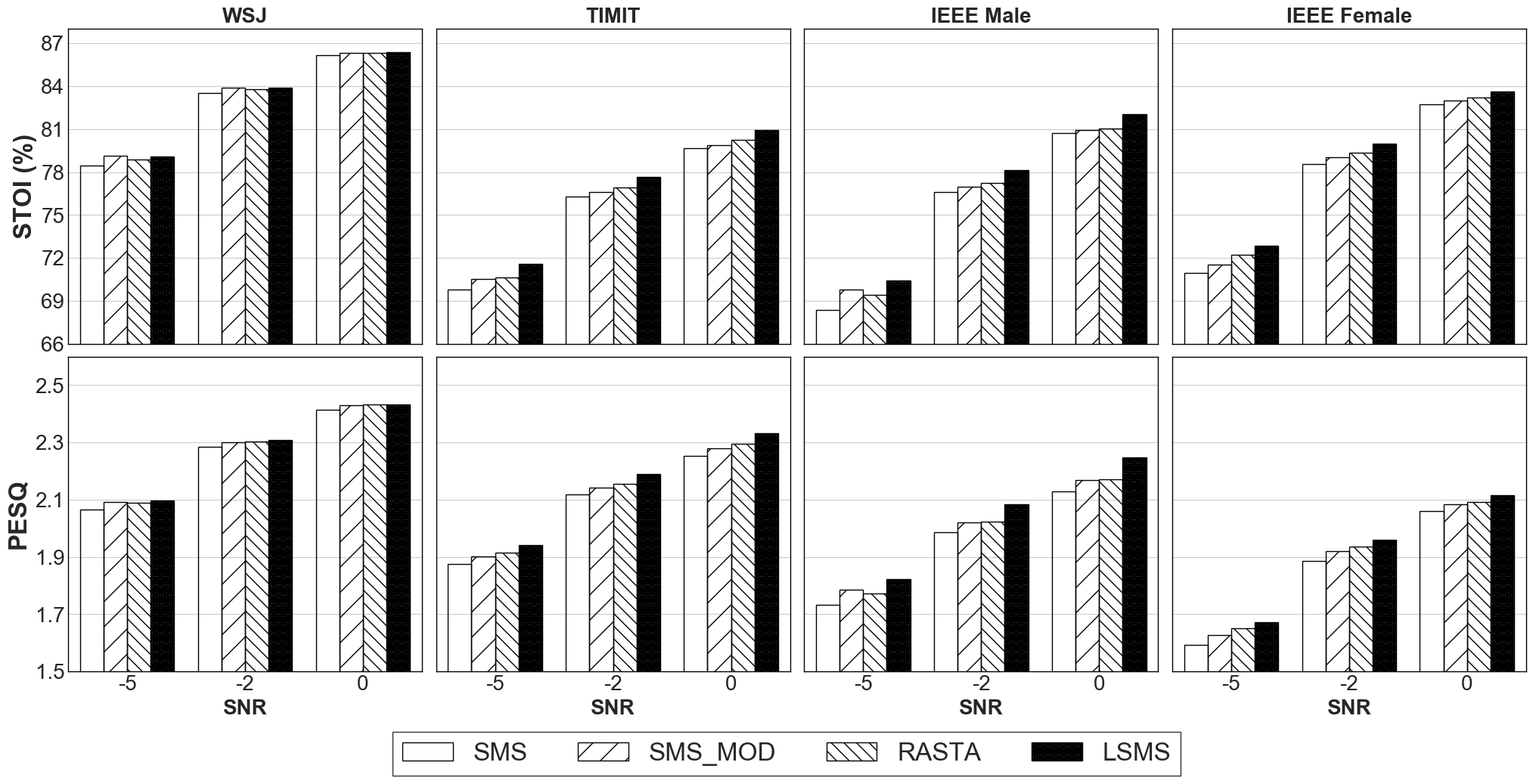}
\caption{STOI and PESQ comparisons between the baseline, modified loss, LSMS and RASTA on WSJ.}
\end{figure*}

\subsection{Channel Normalization}
We have discussed in Section IV that removing the corpus channel can be helpful in improving cross-corpus generalization. We evaluate the following channel normalization techniques in this study.
\subsubsection{Log-Spectral Mean Subtraction}
Given a noisy utterance $\bm{y}$, the channel can be estimated by taking the average of log-spectra over all the frames in the utterance
\begin{equation}
\text{log}|\widehat{V}(f)| = \frac{1}{T} \sum \limits_{t=0}^{T}\text{log}|Y(t, f)|
\end{equation}
The channel normalized log-spectrum is defined as
\begin{equation}
\text{log}|Y^{\prime}(t, f)| = \text{log}|Y(t, f)| -  \text{log}|\widehat{V}(f)|
\end{equation}
We use log$|Y^{\prime}(t, f)|$ as the input feature in this case. Note that estimating the channel using noisy utterances may not be as accurate as  using clean utterances because noise and speech in the data are likely to be recorded in different conditions and using different kinds of devices. Nevertheless, it can give a good approximate for the frequency bins dominated by speech. We add a small positive constant $\epsilon$ before applying the logarithm operator. 
\subsubsection{RASTA Filter}
The RASTA filter has been shown to attenuate the channel effects and improve the generalization of ASR systems \cite{murveit1992reduced}. The RASTA filter is applied over log-spectral magnitude and is given by
\begin{equation}
\begin{split}
\text{log}|Y^{\prime}(t, f)| = \text{log}|Y(t, f)| - \text{log}|Y(t-1, f)| \\ +\ C \cdot \text{log}|Y^{\prime}(t-1, f)|
\end{split}
\end{equation}
 where $C$ is a parameter that is set to $0.97$.
 
 \begin{table}[!b]
\centering
\caption{Different corpus sizes used in this study.}
\centering
\begin{adjustbox}{width=0.95\columnwidth}
\begin{tabular}{|c|c|c|c|c|c|}
\hline
Corpus & WSJ & VoxCeleb2 & LibriClean & LibriOther & LibriAll \\
\hline
\# of speakers & 77 & 5994 & 921 & 1166 & 2087 \\
\hline
\# of utterances & 6385 & 1092009 & 104014 & 148688 & 252702 \\
\hline
\# of hours & 12 & 2318 & 360 & 500 & 860 \\
\hline
\end{tabular}
\end{adjustbox}
\end{table} 

\begin{table}[!b]
\centering
\caption{Learning rate schedule. $E$ denotes the maximum number of epochs of training.}
\centering
\begin{adjustbox}{width=0.85\columnwidth}
\begin{tabular}{|c|c|c|c|}
\hline
Epoch & 1 to 0.6$E$ & (0.6$E$  + 1) to 0.9$E$ & (0.9$E$  + 1) to $E$  \\
\hline
Learning rate & 0.0002 & 0.0001 & 0.00005 \\
\hline
\end{tabular}
\end{adjustbox}
\end{table}

\subsection{Training Corpus}
We evaluate following corpora to understand cross-corpus generalization behavior.
\subsubsection{WSJ}

We use the WSJ0-SI-84 corpus as the baseline since this corpus has been used in past to train speaker- and noise-independent models \cite{chen2017long, tan2018gated1, pandey2019new}.

\subsubsection{VoxCeleb2}
The VoxCeleb2 corpus is promising for cross-corpus generalization because of the following reasons. First, it is very large with around 1.1 million utterances of $6000$ speakers. Second, it is extracted from YouTube therefore it has the potential of generalizing to different channels as the uploaded videos on YouTube are usually recorded in different conditions and using different devices.

\subsubsection{LibriSpeech}
LibriSpeech is a corpus derived from read audiobooks from the LibriVox project. It contains around 0.25 million utterances of 2.1k speakers. It is promising for cross-corpus generalization because the English utterances are spoken by different volunteers across the globe. This implies that the utterances recorded by different volunteers are typically over different channels. 

We have evaluated three different versions of LibriSpeech; LibriClean, LibriOther, and LibriAll. LibriClean contains relatively clean utterances compared to LibriOther. LibriAll is the combination of both LibriClean and LibriOther. We list different corpora in terms of their size in Table II.

\subsection{Frame Shift}
In short-time processing of speech, a frame shift equal to the half of frame size typically is used, and overlap-and-add is used during final reconstruction in the time domain. However, when frame shift is smaller, there will be multiple predictions ($\textgreater$2) of a single T-F unit from the neighboring frames. This leads to averaging the multiple predictions of a sample in the overlap-and-add stage. We find that the simple idea of using a smaller frame shift leads to a significant improvement in cross-corpus generalization. We fix the frame size to $32$ ms and evaluate frame shifts of \{$16$ ms, $8$ ms, $4$ ms, $2$ ms\}.

\begin{figure*}[!t]
\centering
\includegraphics[width=0.75\textwidth]{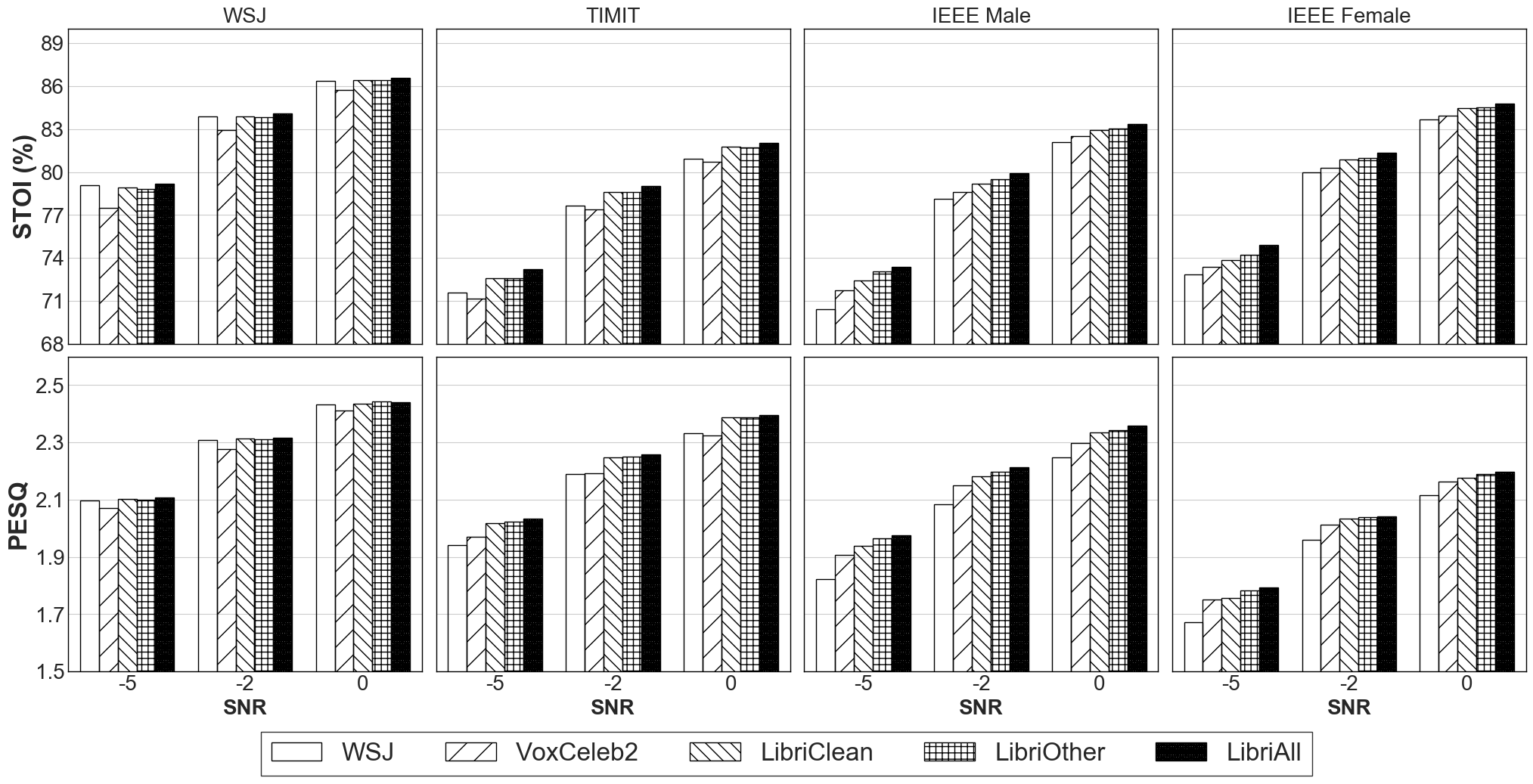}
\caption{STOI and PESQ comparisons between different training corpora with the frame shift of $16$ ms.}
\end{figure*}

\section{Experimental Settings}
\subsection{Data Preparation}
We train corpus dependent models on WSJ, TIMIT, IEEE Male, and IEEE Female corpora. Corpus independent models are trained on WSJ, VoxCeleb2, LibriClean, LibriOther, and LibriAll. For training, we use all $4620$ utterances of the TIMIT corpus and $576$ random utterances out of $720$ of IEEE Male and IEEE Female. All the clean utterances are resampled to $16$kHz. For WSJ training utterances, we remove all the frames in the beginning and end that are not within $20$ dB of the maximum frame energy.

Noisy utterances are created during the training time by randomly adding noise segments to all the utterances in a batch. For training noises, we use $10000$ non-speech sounds from a sound effect library (\url{www.sound-ideas.com}) as in \cite{chen2016large}. For each utterance, we cut a random segment of $4$ seconds if the utterance is longer than $4$ seconds. A random noise segment is added to the utterance at a random SNR in \{$-5$ dB, $-4$ dB, $-3$ dB, $-2$ dB, $-1$ dB, $0$ dB\}. For a corpus containing less than $100000$ utterances, an epoch is defined as when the model has seen around $100000$ utterances. This corresponds to $174$, $22$ and $16$ noisy utterances per clean utterance in one epoch of IEEE, TIMIT, and WSJ respectively.

The WSJ test set consists of $150$ utterances of $6$ speakers not included in WSJ training. The TIMIT test set consists of 192 utterances from the core test set. The IEEE Male and IEEE Female test sets both consist of the $144$ clean utterances not included in their training sets. A test set is generated from $4$ different noises: babble, cafeteria, factory and engine, at the SNRs of \{$-5$ dB, $-2$ dB, $0$ dB\}. The babble and cafeteria noises are from Auditec CD (available at \url{http://www.auditec.com}). Factory and engine noises are from Noisex \cite{varga1993assessment}.

All noisy utterance samples are normalized to the range [$-1$, $1$] and corresponding clean utterances are scaled accordingly to maintain an SNR. The frame size of $32$ ms with the Hamming window is used for STFT.

\subsection{Training Methodology} 
The models trained on TIMIT and IEEE use a dropout rate of 0.5 for each layer except for the output. The models are trained for 10 epochs on TIMIT and IEEE, 100 epochs on LibriSpeech, and 20 epochs on VoxCeleb2.

The Adam optimizer \cite{kingma2014adam} is used with a learning rate schedule given in Table III. A batch size of 32 utterances is used. All the utterances that are shorter than the longest utterance in a batch are padded with zero at the end. The loss values computed over the outputs corresponding to zero-padded inputs are ignored.

\begin{figure*}[!t]
\centering
\includegraphics[width=0.75\textwidth]{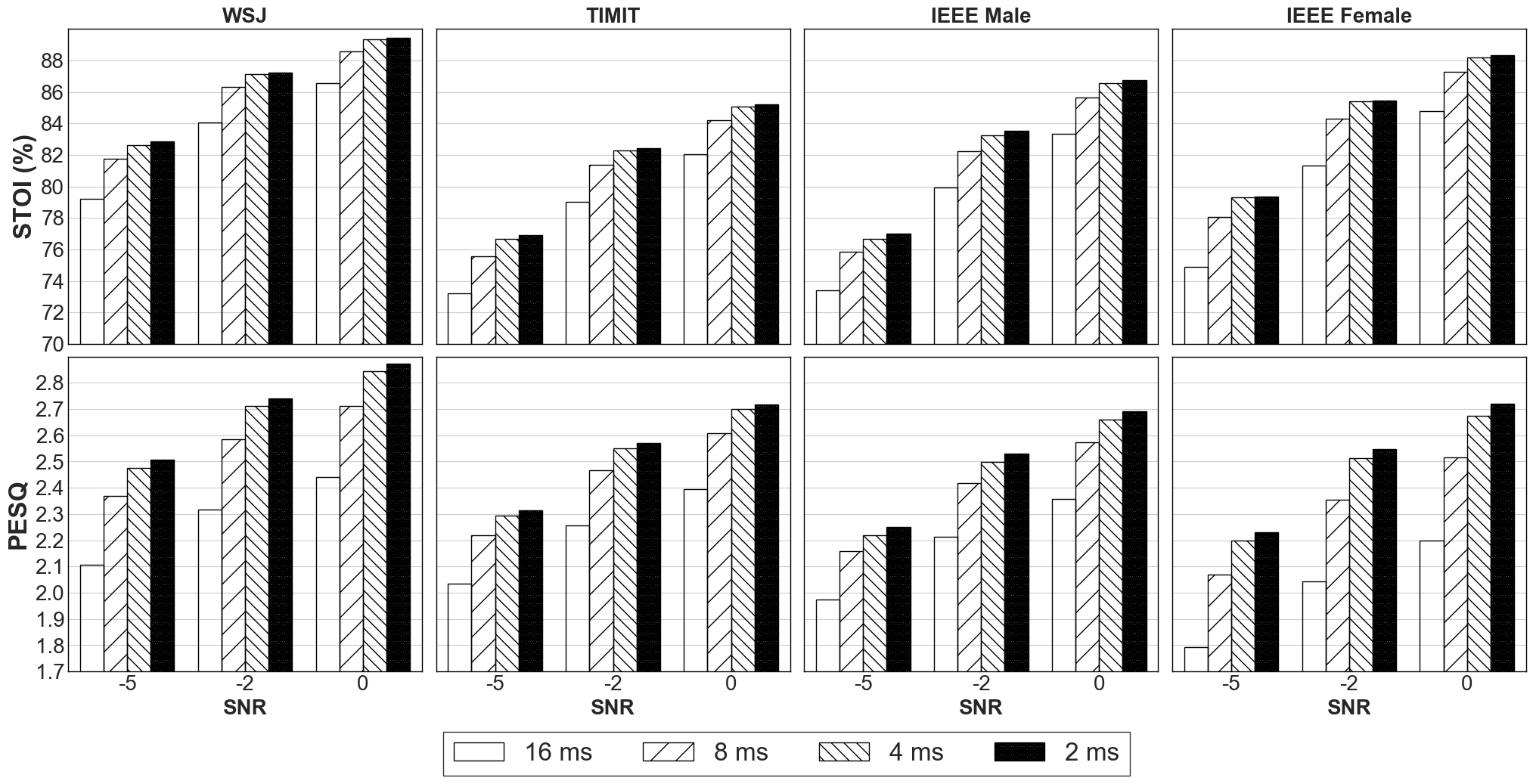}
\caption{STOI and PESQ comparisons between different frame shifts on LibriAll.}
\end{figure*}

\subsection{Evaluation Metrics}
In our experiments, models are evaluated using STOI \cite{taal2011algorithm} and PESQ \cite{rix2001perceptual}, which represent the standard metrics for speech enhancement. STOI has a typical value range from $0$ to $1$, which can be roughly interpreted as percent correct. PESQ values range from $-0.5$ to $4.5$.

\subsection{Baseline}
For the baseline, we train the BLSTM model on WSJ using the loss function given in Equation 8. STFT magnitude is used as the feature with the channel normalization in Equation 22 but applied to STFT magnitude instead of log magnitude. We call this model SMS, standing for spectral mean subtraction (in Fig. 5 and Table IV).

\section{Results and Discussions}
First, we evaluate the modified loss function (Section V.A) and two channel normalization methods (Section V.B) and compare them with the baseline model. The models are trained on the WSJ corpus with a frame shift of $16$ ms. We denote the baseline with SMS and the model with modified loss as SMS\_MOD. Average STOI and PESQ over all the four test noises and at SNRs of $-5$ dB, $-2$ dB, and $0$ dB are plotted in Fig. 5. 

We observe that SMS\_MOD is consistently better than SMS. The improvement is maximum at $-5$ dB for all the corpora. The maximum improvement is observed for the IEEE Male corpus. The objective scores indicate that training a model using a loss over all the T-F units leads to overfitting on the corpus. Using a loss computed over only high energy T-F units can achieve better generalization. All the following models trained in this study, except for SMS, will use the modified loss function.

The objective scores for two normalization schemes suggest that LSMS and RASTA both are better than SMS and SMS\_MOD for all untrained corpora. LSMS is consistently better than RASTA for all the corpora and at all SNR conditions.

\begin{figure*}[!t]
\centering
\includegraphics[width=0.75\textwidth]{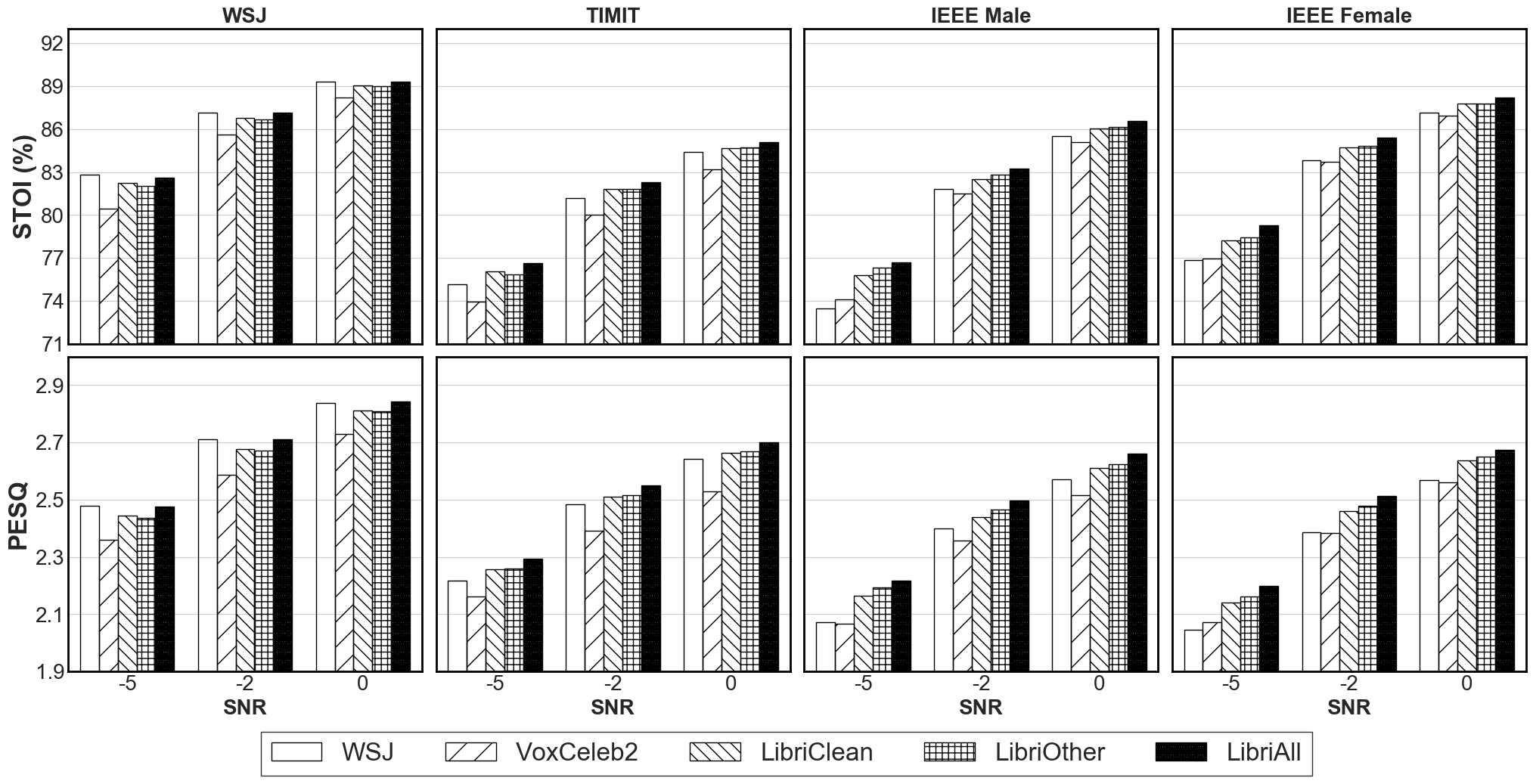}
\caption{STOI and PESQ comparisons between different training corpora with the frame shift of $4$ ms.}
\end{figure*}

Next, we examine different training corpora on $4$ test noises. The models are trained using LSMS with a frame shift of $16$ ms. The average STOI and PESQ over four test noises are plotted in Fig. 6. A general trend for STOI and PESQ scores are $\text{LibriAll} > \text{LibriOther} > \text{LibriCLean} > \text{VoxCeleb2} > \text{WSJ}$, except for TIMIT where VoxCeleb2 is worse than WSJ. 

A key observation from the corpora comparisons is that the corpus content is important to achieve better generalization but not the size of the corpus. A corpus with multiple possible channels sources, LibriAll, is very effective for generalization. However, a similar corpus VoxCeleb2 containing $4.3$ times more utterances is not as effective. This observation is further supported by the fact that no dramatic performance differences exist between LibriClean ($104014$ utterances), LibriOther ($148688$ utterances) and LibriAll ($252702$ utterances), all of which contain utterances from the LibriSpeech corpus.

Perhaps surprisingly, VoxCeleb2 is not able to obtain good generalization. This might be due to the types of utterances in VoxCeleb2. Most of the utterances include some sort of reverberation, cross-talk or background noise. Hence, it may not be very suitable to be employed for the enhancement of utterances from clean corpora. More research is needed to explain the cross-corpus generalization behavior of VoxCeleb2.

Further, we compare models trained with different frame shifts. We compare frame shifts from \{$16$ ms, $8$ ms, $4$ ms, $2$ ms\}. All the models are trained on LibriAll using LSMS with a frame size of 32 ms. Average STOI and PESQ scores are plotted in Fig. 7. We can observe a clear improvement in the objective scores when moving from $16$ ms to $8$ ms, and from $8$ ms to $4$ ms. However, the performances for $4$ ms and $2$ ms are very similar, suggesting the diminishing effect from reducing frame shift. Note that similar performance improvements are obtained using all the training corpora, suggesting that using small frame shift is an effective technique applicable to all training corpora. The performance is also improved on the trained corpus, WSJ in this case, when trained using smaller frame shifts. This is an important observation because getting an improvement on the trained corpus does not necessarily result in an improvement over untrained corpora as we have reported in Table I.  

We also compare all the training corpora using a smaller frame shift of $4$ ms and the results are plotted in Fig. 8. We obtain the same performance trend as using the frame shift of $16$ ms. This implies that using smaller frame shift and better training corpora are two independent techniques for improving cross-corpus generalization.
\begin{table}[!b]
\centering
\caption{Performance improvements on babble noise by gradually incorporating different techniques proposed in this study.}
\centering
\begin{adjustbox}{width=0.95\columnwidth}
\begin{tabular}{|c|c|cc|cc|cc|cc|}
\hline
\multicolumn{2}{|c|}{ Test Corpus } & \multicolumn{2}{c|}{ WSJ } & \multicolumn{2}{c|}{ TIMIT } & \multicolumn{2}{c|}{ IEEE Male } & \multicolumn{2}{c|}{ IEEE Female } \\
\hline
\multicolumn{2}{|c|}{ Test SNR } & -5 dB & -2 dB & -5 dB & -2 dB & -5 dB & -2 dB & -5 dB & -2 dB \\
\hline
\hline
\multirow{7}{*}{ \rotatebox{90}{STOI (\%)} } & Mixture & 58.6 & 65.5 & 54.0 & 60.9 & 55.0 & 62.3 & 55.5 & 62.9 \\
\cline{2-10}
& Baseline & 77.4 & 83.0 & 64.7 & 73.3 & 60.4 & 74.0 & 62.5 & 73.5 \\
& + Modified loss & 78.3 & 83.5 & 65.7 & 74.3 & 64.8 & 75.1 & 63.8 & 75.2 \\
& + LSMS & 78.6 & 83.6 & 68.4 & 76.4 & 64.4 & 76.6 & 66.0 & 76.7 \\
& + frame shift 4 ms & \textbf{82.8} & \textbf{87.5} & 71.9 & 79.9 & 66.2 & 80.8 & 69.5 & 81.1 \\
& + LibriAll & 82.4 & 87.3 & \textbf{75.1} & \textbf{82.1} & \textbf{74.3} & \textbf{83.2} & \textbf{74.8} & \textbf{84.3} \\
\cline{2-10}
& Same Corpus & - & - & \textbf{73.5} & \textbf{80.7} & \textbf{77.9} & \textbf{82.6} & \textbf{75.9} & \textbf{83.2} \\
\hline
\hline
\multirow{7}{*}{  \rotatebox{90}{PESQ }} & Mixture & 1.54 & 1.69 & 1.46 & 1.63 & 1.46 & 1.63 & 1.12 & 1.32 \\
\cline{2-10}
& Baseline & 1.97 & 2.22 & 1.70 & 2.00 & 1.52 & 1.89 & 1.26 & 1.66 \\
& + Modified loss & 2.00 & 2.23 & 1.73 & 2.04 & 1.63 & 1.92 & 1.31 & 1.74 \\
& + LSMS & 2.02 & 2.25 & 1.82 & 2.12 & 1.64 & 2.00 & 1.39 & 1.81 \\
& + 4 ms frame shift & \textbf{2.45} & \textbf{2.72} & 2.09 & 2.43 & 1.8 & 2.33 & 1.67 & 2.22 \\
& + LibriAll & 2.43 & 2.70 & \textbf{2.20} & \textbf{2.52} & \textbf{2.11} & \textbf{2.47} & \textbf{1.94} & \textbf{2.41} \\
\cline{2-10}
& Same Corpus & - & - & \textbf{2.12} & \textbf{2.42} & \textbf{2.14} & \textbf{2.38} & \textbf{2.03} & \textbf{2.40} \\
\hline
\end{tabular}
\end{adjustbox}
\end{table}

Furthermore, we report results on babble noise when different techniques to improve channel generalization are gradually incorporated into the baseline model. The results are given in Table IV. The bold scores in the last row of STOI and PESQ, Same Corpus (trained corpus), provide the scores obtained by training a model on the same corpus as the test corpus. Note that the results on the trained corpora, TIMIT and IEEE, represent benchmarks where the number of unique training utterances is small. IEEE corpora have only $576$ training utterances and TIMIT has $4620$ utterances in which many speakers speak the same set of sentences. A good model should be able to match the scores obtained using Same Corpus.

We observe that the most effective approach is the use of LibriAll that improves STOI at $-5$ dB by $3.2\%$ on TIMIT, $8.1\%$ on IEEE Male, and $5.3\%$ on IEEE Female while obtaining similar performance on WSJ as to that obtained by training on WSJ. Similarly, smaller frame shift is also very effective as it improves STOI at $-5$ dB by $3.5\%$ on TIMIT, $1.8\%$ on IEEE Male, and $3.5\%$ on IEEE Female.
\begin{table}[!b]
\centering
\caption{Performance improvements on reverberant speech mixed with babble noise by gradually incorporating different techniques.}
\centering
\begin{adjustbox}{width=0.95\columnwidth}
\begin{tabular}{|c|c|cc|cc|cc|cc|}
\hline
\multicolumn{2}{|c|}{ Test Corpus } & \multicolumn{2}{c|}{ WSJ } & \multicolumn{2}{c|}{ TIMIT } & \multicolumn{2}{c|}{ IEEE Male } & \multicolumn{2}{c|}{ IEEE Female } \\
\hline
\multicolumn{2}{|c|}{ Test SNR } & -5 dB & -2 dB & -5 dB & - 2dB & -5 dB & - 2 dB & -5 dB & - 2 dB \\
\hline
\multirow{7}{*}{ \rotatebox{90}{STOI (\%)} } & Mixture & 53.26 & 57.1 & 50.07 & 54.67 & 53.98 & 59.27 & 52.98 & 57.75 \\
\cline{2-10}
& Baseline & 65.1 & 68.4 & 54.3 & 60.4 & 57.6 & 66.4 & 54.9 & 61.1 \\
& +Modified loss & 64.2 & 67.8 & 54.9 & 59.6 & 56.8 & 65.2 & 55.0 & 61.9 \\
& +LSMS & 67.5 & 71.0 & 57.8 & 64.8 & 57.3 & 67.6 & 56.7 & 63.3 \\
& + frame shift 4ms & 69.8 & 73.3 & 59.7 & 65.7 & 61.6 & 70.1 & 57.4 & 65.3 \\
& +LibriAll & \textbf{70.8} & \textbf{73.5} & \textbf{61.4} & \textbf{68.4} & \textbf{63.9} & \textbf{73.2} & \textbf{63.7} & \textbf{70.3} \\
\cline{2-10}
& Same Corpus & - & - & \textbf{62.8} & \textbf{68.5} & \textbf{65.2} & \textbf{71.5} & \textbf{65.3} & \textbf{71.4} \\
\hline
\hline
\multirow{7}{*}{ \rotatebox{90}{PESQ} } & Mixture & 1.40 & 1.53 & 1.36 & 1.49 & 1.39 & 1.56 & 1.03 & 1.22 \\
\cline{2-10}
& Baseline & 1.65 & 1.87 & 1.45 & 1.67 & 1.45 & 1.73 & 1.09 & 1.35 \\
& +Modified loss & 1.61 & 1.82 & 1.44 & 1.65 & 1.45 & 1.73 & 1.09 & 1.36 \\
& +LSMS & 1.80 & 2.02 & 1.58 & 1.88 & 1.49 & 1.88 & 1.18 & 1.50 \\
& + frame shift 4ms & 1.99 & 2.23 & 1.66 & 1.97 & 1.64 & 1.97 & 1.25 & 1.67 \\
& +LibriAll & \textbf{2.09} & \textbf{2.28} & \textbf{1.77} & \textbf{2.09} & \textbf{1.78} & \textbf{2.15} & \textbf{1.57} & \textbf{1.94} \\
\cline{2-10}
& Same Corpus & - & - & \textbf{1.84} & \textbf{2.11} & \textbf{1.77} & \textbf{2.05} & \textbf{1.66} & \textbf{2.02} \\
\hline
\end{tabular}
\end{adjustbox}
\end{table}

All the proposed techniques are trained and evaluated on corpora with negligible room reverberation.  Speech enhancement in the presence of both reverberation and background noise at low SNRs, such as $-5 $ dB, is an extremely difficult problem, and would require training with noisy-reverberant utterances \cite{zhao2018two}. To examine the generality of the proposed techniques, we further evaluate on noisy-reverberant speech data. To create reverberant utterances, we utilize real room impulse responses (RIRs) in  \cite{hummersone2010dynamic}. We use all $74$ RIRs corresponding to the room with the reverberation time of $0.32$ seconds. A given clean utterance is convolved with a randomly picked RIR, and is followed by noise addition. The results are reported in Table V, where anechoic speech is considered the reference signal in the evaluation. Note that the models already trained without reverberation are tested without retraining, and hence it is expected that the amounts of improvement are lower than those in Table IV. However, we observe a similar trend of cross-corpus generalization, except for the modified loss which is worse than the baseline. The model trained on LibriAll using LSMS with a frame shift of $4$ ms performs the best in this case as well.

\section{Concluding Remarks}
This work reveals robustness problem with deep learning based speech enhancement algorithms. We have shown that a model trained on a given corpus fails to generalize to utterances from an untrained corpus. The problem is more severe at low SNR levels, where speech enhancement is actually more needed. We have established that the cross-corpus generalization issue is mainly due to the channel mismatch between a trained and untrained corpus.

We have examined traditional channel normalization methods and found that they improve performance on untrained corpora, but improvement is limited, and hence other techniques need to be developed to further improve generalization. 

We have proposed two effective methods to significantly improve cross-corpus generalization. The first technique is to use a corpus obtained using crowd-sourced audio recordings such as LibriSpeech and VoxCeleb. We found LibriSpeech to be significantly better than VoxCeleb. The second technique is the use of a smaller frame shift in STFT and ISTFT layers.

Further research is needed to evaluate the effectiveness of LibriSpeech and smaller frame shift for complex-domain and time-domain speech enhancement models. The behavior of VoxCeleb, which is found to be not very effective for generalization, needs to be further explored for a better understanding of cross-corpus generalization.

\ifCLASSOPTIONcaptionsoff
  \newpage
\fi

\bibliographystyle{IEEEtran}
\bibliography{mybib}

\vspace{-1.1cm}
\begin{IEEEbiography}[{\includegraphics[width=2.5cm,height=3cm]{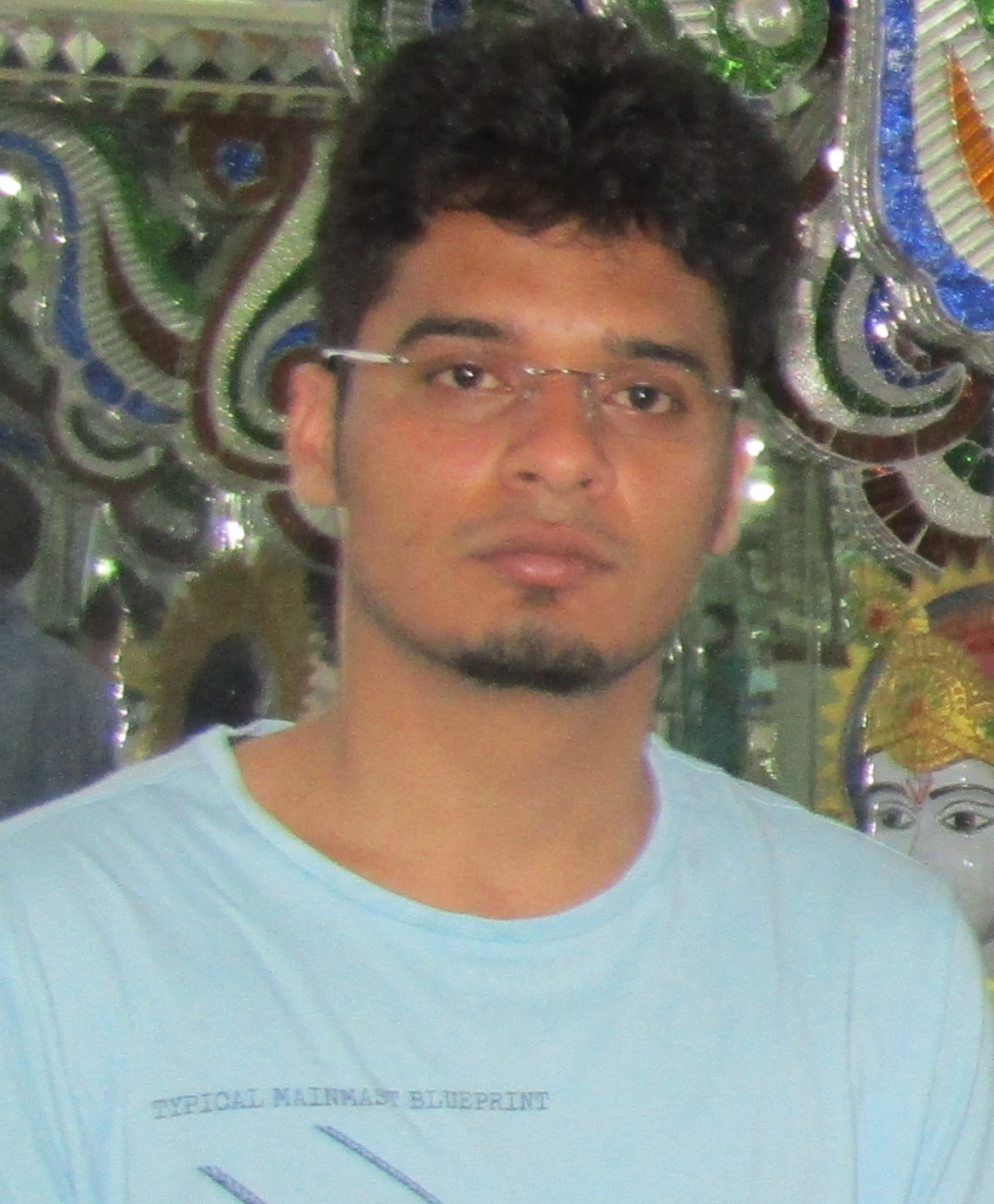}}]{Ashuosh Pandey} received his B.Tech degree in Electronics and communication engineering from Indian Institute of Technology, Guwahati, India, in 2011. He is currently pursuing his Ph.D. degree at The Ohio State University. He is interested in speech separation and deep learning.
\end{IEEEbiography}

\end{document}